# Ferromagnetic-like behavior of $Bi_{0.9}La_{0.1}FeO_3$ − KBr nanocomposites

**Dmitry V. Karpinsky[1], Olena M. Fesenko[2], Maxim V. Silibin[3,4], Sergei V. Dubkov[3], Mykola Chaika[2], Andriy Yaremkevich[2], Anna Lukowiak[5], Yuriy Gerasymchuk[5], Wiesław Stręk[5], Andrius Pakalniškis[6], Ramunas Skaudzius[6], Aivaras Kareiva[6], Yevhen M. Fomichov[7], Vladimir V. Shvartsman[8] Sergei V. Kalinin[9], Nicholas V. Morozovsky[2], and Anna N. Morozovska[2*]**

[1] Scientific-Practical Materials Research Centre of NAS of Belarus, 220072 Minsk, Belarus

[2] Institute of Physics, NAS of Ukraine, 46, pr. Nauky, 03028 Kyiv, Ukraine[4]

[3] National Research University of Electronic Technology "MIET", 124498 Moscow, Russia

[4] Institute for Bionic Technologies and Engineering, I.M. Sechenov First Moscow State Medical University, Moscow 119991, Russia

[5] Institute of Low Temperature and Structure Research, PAS, Wroclaw, 50-422, Poland

[6] Institute of Chemistry, Vilnius University, Naugarduko 24, Vilnius, LT-03225, Lithuania

[7] Charles University in Prague, Faculty of Mathematics and Physics, V Holešovičkach 2, Prague 8, 180 00, Czech Republic

[8] Institute for Materials Science and Center for Nanointegration Duisburg-Essen (CENIDE), University of Duisburg-Essen, 45141, Essen, Germany

[9] The Center for Nanophase Materials Sciences, Oak Ridge National Laboratory, Oak Ridge, TN 37831

**Abstract**

We studied magnetostatic response of the $Bi_{0.9}La_{0.1}FeO_3$ - KBr composites (BLFO-KBr) consisting of nanosized (≈100 nm) ferrite $Bi_{0.9}La_{0.1}FeO_3$ (BLFO) conjugated with fine grinded ionic conducting KBr. When the fraction of KBr is rather small (less than 15 wt %) the magnetic response of the composite is very weak and similar to that observed for the BLFO (pure KBr matrix without $Bi_{1-x}La_xFeO_3$ has no magnetic response as anticipated). However, when the fraction of KBr increases above 15%, the magnetic response of the composite changes substantially and field dependence of magnetization reveals ferromagnetic-like hysteresis loop with a remanent magnetization about 0.14 emu/g and coercive field about 1.8 Tesla (at room temperature). Nothing similar to the ferromagnetic-like hysteresis loop can be observed in BLFO ceramics, which magnetization quasi-linearly increases with magnetic field.

---

[*] Corresponding author: anna.n.morozovska@gmail.com

Different physical mechanisms were considered to explain the unusual experimental results for BLFO-KBr nanocomposites, but only those among them, which are highly sensitive to the interaction of antiferromagnetic $Bi_{0.9}La_{0.1}FeO_3$ with ionic conductor KBr, can be relevant. An appropriate mechanism turned out to be ferro-magneto-ionic coupling.

## I. INTRODUCTION

### A. Magnetic, structural and electrophysical properties of pristine $BiFeO_3$

Multiferroics are ideal systems for fundamental studies of couplings among the order parameters of different nature, e.g. ferroelectric (**FE**) polarization, structural rotational antiferrodistortion (**AFD**), ferromagnetic (**FM**) and antiferromagnetic (**AFM**) long-range order parameters [1, 2, 3, 4, 5, 6]. Magnetoelectric (**ME**) coupling is especially important for the most of multiferroic fundamental studies and applications [1-5].

Bismuth ferrite $BiFeO_3$ (**BFO**) is the unique multiferroic with the large FE polarization (more than 60 μC/cm$^2$ at RT) and AFM order coexisting up to room (RT) and elevated temperatures [7, 8]. BFO exhibits unusual electrophysical properties, such as conduction and magnetotransport enhancement at domain walls [9, 10, 11, 12, 13, 14]. Specifically, bulk BFO exhibits AFD long-range order at temperatures below 1200 K; it is FE below Curie temperature $T_C \approx 1145$ K and is AFM below Neel temperature $T_N \approx 645$ K [15]. It is known that pristine bulk BFO is characterized by a cycloidal modulation of the magnetization superimposed on the AFM G-type magnetic structure [3]. Although the linear magnetoelectric effect is symmetry forbidden on a macroscopic scale, the Dzyaloshinskii-Moriya mechanism can be used locally in BFO to achieve an electric switching of the spin cycloid [16, 17, 18, 19].

### B. Magnetic, structural and electrophysical properties of $Bi_{1-x}La_xFeO_3$

Chemical doping of pristine BFO with lanthanum La ions having ionic radius (1.16Å [20]) similar to that of bismuth Bi ions (1.17Å [20]) causes a structural transition from the polar rhombohedral phase to the anti-polar orthorhombic phase which is accompanied with a slight decrease in the unit cell volume [21, 22, 23]. The structural transition is associated with significant changes in the unit cell parameters, namely, the modification of the Fe–O chemical bond lengths (a shortening of the bond length between the iron ions and apical oxygen ions and an elongation of the chemical bond between the iron ions and the oxygen ions located in the basal *ab* plane), an increase in the Fe–O–Fe angles, decrease in the oxygen octahedra rotations and tilts etc. The mentioned changes of the structural parameters lead to a reduction of the polar displacement of the ions and to significant change of the magnetic properties of the compounds near the rhombohedral-orthorhombic phase boundary. The diffraction data obtained at RT for the compounds $Bi_{1-x}La_xFeO_3$ (**BLFO**) indicate a single phase rhombohedral state for the compounds with x < 0.16, two phase structural state assuming a coexistence of the polar rhombohedral and the anti-polar orthorhombic phases in the concentration range 0.16 < x < 0.19; further increase in the dopant content leads to a stabilization of single phase orthorhombic state (incommensurately modulated anti-polar orthorhombic phase, anti-polar or non-polar orthorhombic phase) depending on the La content as described elsewhere [24, 25, 26].

Sol-gel synthesis method used to prepare BLFO solid solution leads to a formation of oxygen stoichiometric compounds those magnetic properties are governed by the exchange interactions between Fe ions being in 3+ oxidation state. Increase in the concentration of the lanthanum La ions leads to a gradual reduction in the critical magnetic field associated with a disruption of the spatially modulated spin structure and the compound with lanthanum content $x \sim 0.17$ is characterized by weak ferromagnetic structure at RT assuming complete disruption of the spatially modulated structure. The structural transition from the rhombohedral to the orthorhombic phase is accompanied by a significant increase of the remanent magnetization and coercive field, which is observed for the compounds having dominant or single phase orthorhombic structure.

It should be noted that transport properties of the BLFO compounds are also strongly dependent on the La concentration, and the conductivity of these solid solutions gradually increases with the dopant content [27]. Impedance measurements have shown mainly p-type conductivity of the compounds having single phase rhombohedral structure which significantly increases with temperature [27].

This work studies magnetostatic response of the newly synthesized $Bi_{0.9}La_{0.1}FeO_3$ - KBr composites (BLFO-KBr) consisting of nanosized (≈100 nm) ferrite $Bi_{0.9}La_{0.1}FeO_3$ (BLFO) conjugated with fine grinded ionic conducting KBr. We revealed that when the fraction of KBr increases above 15%, it demonstrates ferromagnetic-like hysteresis loop with a remanent magnetization about 0.14 emu/g and coercive field about 1.8 Tesla (at room temperature). To the best of our knowledge nothing similar was reported previously in the literature. Different physical mechanisms were considered to explain the unusual experimental results for BLFO-KBr nanocomposites. The original part of the manuscript is structured as follows. Magnetostatic and electrophysical properties of BLFO–KBr nano-composites are discussed in **section II**. Theoretical explanation of the ferromagetism in BLFO–KBr nano-composites is given in **section III**. **Section IV** is a brief summary. Auxiliary experimental results and samples characterization are given in the Supplement.

# II. MAGNETIC AND ELECTROPHYSICAL PROPERTIES OF BLFO–KBr NANO-COMPOSITE

## A. Nanocomposite preparation and characterization

The samples of nanograined BLFO ceramics have been prepared by the aqueous sol-gel method using hydrates $Bi(NO_3)_3 \cdot 5H_2O$, $Fe(NO_3)_3 \cdot 9H_2O$ and lanthanum nitrate as starting materials. The prepared sol-gel samples were heated at 800°C for 1.5 h [28, 29]. The composites have been prepared by thorough mechanical mixing of the single phase ferrite ($Bi_{0.9}La_{0.1}FeO_3$) and grinded KBr taken in the next mass ratios – 90:10, 85:15, 80:20, 70:30, 50:50 respectively. The obtained powders have been uniaxially pressed at compacting pressures of 5GPa to obtain tablets with 5mm diameter and 1mm thickness.

The compound $Bi_{0.9}La_{0.1}FeO_3$, which has been mainly used to prepare the composites, is characterized by the grains with average radius of about 100 nm, and the small amount of the grains have the average radius of about 35-40 nm (the SEM image and the histogram of the grain sizes distribution are shown in the **Fig. 1(a)**. SEM measurements performed for the composites have allowed to clarify their morphology and to specify a distribution of the constituent phases. The crystallines having rounded-shaped form are attributed to the KBr phase as determined by EDS analysis carried out for the selected areas of the SEM images [**Fig. 1(b)**]. Characteristic size of the grains attributed to KBr phase is in the range of $0.100 – 10$ μm. The crystallines ascribed to the ferrite phase are characterized by rectangular-shaped form, ferrite grains are mainly agglomerated into clusters having size of about 1-2 microns, while separate grains are characterized by nanometer size as confirmed by SEM images obtained for the composites and the single phase ferrite [compare **Fig. 1(a)** and **1(b)**]. The SEM data have confirmed homogeneous distribution of the constituent phases in the composites which facilitates active chemical processes in the boundary region of both phases.

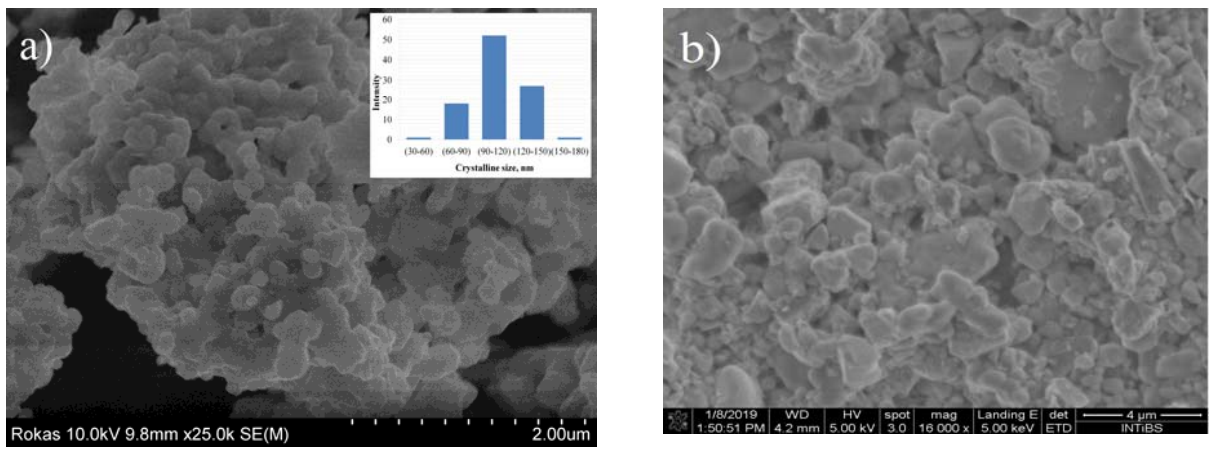

**Figure 1. (a)** SEM image of the compound $Bi_{0.9}La_{0.1}FeO_3$ obtained at room temperature. The inset shows a distribution of crystalline size over the SEM image of the compound. **(b)** SEM image of the composite 50% $Bi_{0.9}La_{0.1}FeO_3$ -50% KBr**.**

The modified sol-gel method synthesis method and preparation conditions [28, 29] used to synthesize nanoscale powder of $Bi_{1-x}La_xFeO_3$ (x =0, 0.05, 0.1) lead to a formation of single phase stoichiometric ferrites utilized to form the composites. The X-ray diffraction measurements performed for the ferrite compounds have confirmed their single phase rhombohedral structure, the XRD patterns recorded at RT have been successfully refined assuming non-centrosymmetric rhombohedral space group *R3c*.

Under standard conditions, potassium bromide (KBr) is a white nonmagnetic crystalline powder which exhibits high ionic conductivity. Phase purity of the commercial KBr powder has been confirmed by the XRD measurements performed by the authors, the obtained results are in accordance with the structural data provided by the producer (see **Supplement, Fig. S1**). The XRD patterns obtained for the composites have confirmed a coexistence of two constituents having the phase ratios in accordance with the mentioned chemical formulas. The diffraction peaks attributed to the constituents do not show any modification of the reflections positions, their width, asymmetry etc. as compared to the reflections observed on the diffraction patterns obtained separately for BLFO and KBr (see **Supplement, Fig. S1**). Phase purity of the constituents assumes that the magnetic properties of the composites are determined only by the ferrite component.

For comparison with the composite, the temperature dependences of magnetization for the nanograined ferrites $Bi_{1-x}La_xFeO_3$ (x = 0, 0.05 and 0.1) have been measured in the temperature range 300 – 1000 K using MPMS SQUID VSM magnetometer (**Supplement, Fig. S2**). The isothermal magnetization measurements have been done at room temperature and at T = 5K (**Supplement, Fig. S3**). The obtained results demonstrate a quasi-linear behavior of the magnetization dependences (see **Supplement, Fig. S3**). An increase of the magnetization observed for lightly doped compounds in strong magnetic fields is attributed to a partial disruption of the cycloidal modulation superimposed on the G-type antiferromagnetic structure [30, 31]. The very small values of the remanent magnetization (0.015-0.02 emu/g) observed for the lightly-doped compounds mainly depends on the dopant content and is caused by uncompensated spin magnetic moments formed in the surface layer of the crystallites due to disruption of the modulated magnetic structure. The compound $Bi_{0.9}La_{0.1}FeO_3$ having rhombohedral crystal structure and modulated magnetic structure with almost zero remanent magnetization has been selected as the most appropriate compound for formation of the composites to estimate an influence of the constituents on the physical properties of nanocomposites.

## B. Magnetic and electrophysical properties of $(Bi_{0.9}La_{0.1}FeO_3)_x$–$(KBr)_{1-x}$ nano-composites

The isothermal dependences of magnetization (M) on magnetic field (H) of the nanocomposites $(Bi_{0.9}La_{0.1}FeO_3)_x$–$(KBr)_{1-x}$ (shortly **(BLFO)$_x$–(KBr)$_{1-x}$**) are shown in **Fig. 2** for the fractions x = 100% (red symbols), 90% (blue), 85% (green), 80% (violet), 70% (brown) and 50% (black). The M-H-dependences have been obtained in the fields up to 14 T using a magnetometer from Cryogenic Ltd. We have also confirmed experimentally that KBr matrix without ferrite compound has no magnetic response. **Fig. 2,** left inset shows KBr-fraction (1-x)-dependences of the remanent magnetization (**M$_r$**), maximal magnetization (**M$_m$**) at the maximal field $H_m$ = 14 T and the coercive field (**H$_c$**) for (BLFO)$_x$–(KBr)$_{1-x}$ composites.

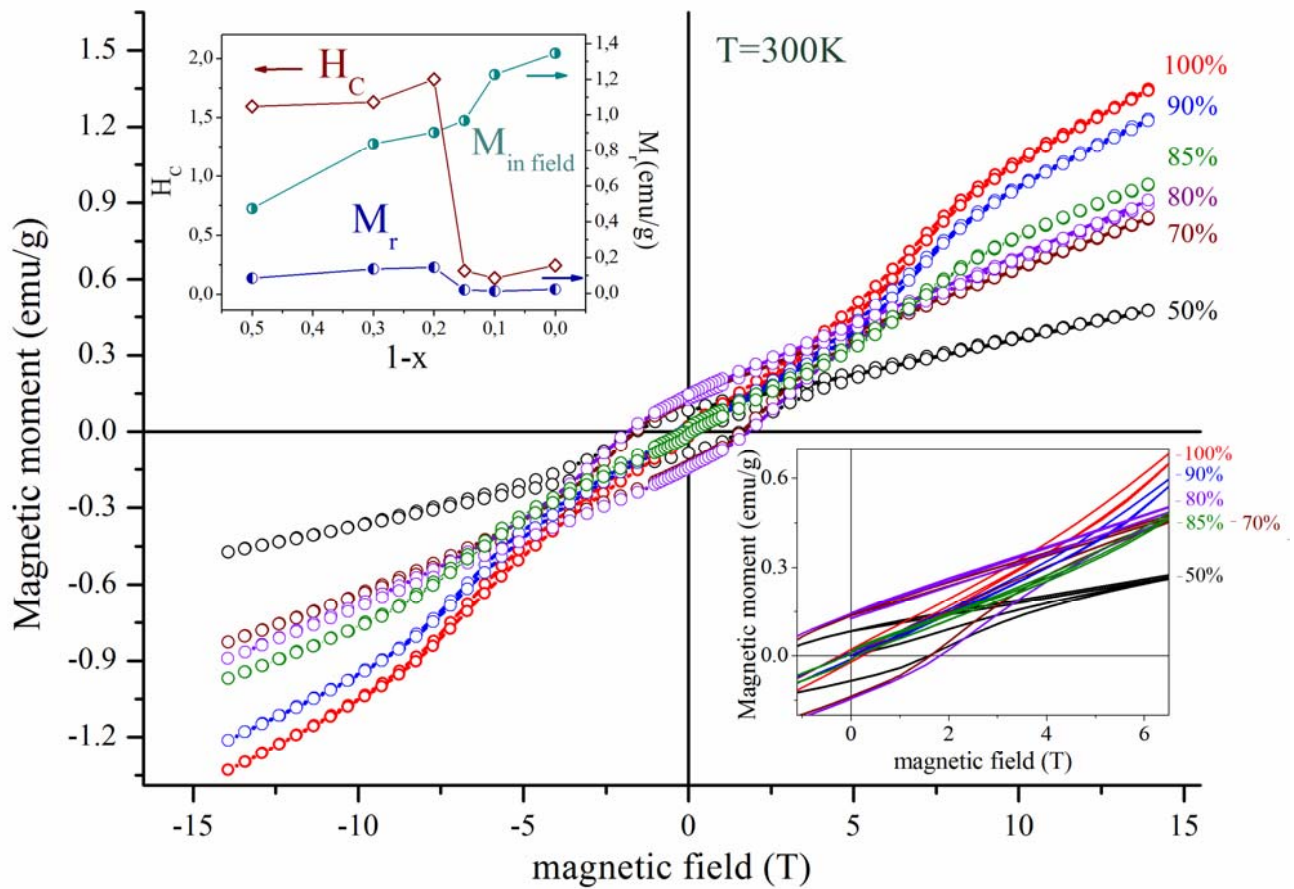

**Figure 2.** Dependences of magnetic moment versus applied magnetic field for the composites (BLFO)$_x$ – (KBr)$_{1-x}$ at RT (the data for composites with the fractions of x = 100% are denoted by red symbols, 90% - blue, 85% - green, 80% - violet, 70% - brown and 50% - by black symbols. Left inset shows KBr fraction (1-x)-dependences of M$_r$, M$_m$ and H$_c$ for the composites; right inset shows enlarged part of the M(H) dependences.

The M-H-dependences of the composites with the fraction of KBr less than 15% are similar to those observed for the compounds without KBr content (see red, blue and green symbols in **Fig. 2**). Further increase in the KBr fraction significantly changes the magnetic properties of the composites (BLFO)$_x$–(KBr)$_{1-x}$, namely leads to the appearance of ferromagnetic-like hysteresis of M-H-loops (see violet, brown and black symbols in **Fig. 2**). Note that the increase of the KBr fraction (1-x) above 15% leads to the step-like decrease of $M_m$, and increase of $M_r$ and $H_c$ values of the composites (see **Fig. 2** insets). The composite containing 20% of KBr is characterized by $M_r \approx 0.15$ emu/g and $H_c \approx 1.8$T at RT (violet symbols in **Fig. 2** and inset). Further increasing in the KBr fraction leads to the gradual decrease of $M_r$, which is mainly associated with a dilution of the magnetically active ferrite component, wherein $H_c$ attributed to the composites remains nearly unchanged (see violet, brown and black symbols in **Fig. 2** and inset). The composite with equal fractions of the constituents (x = 50%) is characterized by the opened magnetization loop specific for ferromagnetic-like materials. Magnetization measurements performed at temperature ~ 5 K do not reveal any notable changes of the composites' magnetic properties, slight increase in magnetization observed at temperature of ~ 5 K can be justified by a reinforcement of the magnetic interactions occurred at low temperatures.

Different physical mechanisms can be responsible for the appearance of RT ferromagnetic-like response of (BLFO)$_x$–(KBr)$_{1-x}$ nanocomposites, where BLFO with La content less that 16% is purely antiferromagnetic in the bulk at temperatures lower than 650 K and KBr is magnetically inactive material. Anyway, the changes observed in the magnetization behavior of the composites can be completely attributed to the changes occurred in the ferrite component. One of the most plausible models, describing the changes of the magnetic properties observed for the composites (BLFO)$_x$–(KBr)$_{1-x}$ depending on the KBr content assumes intensive chemical processes in the vicinity of ferrite particle - salt interface occurred during the composites formation. As a result of the redox reactions the Fe ions near the complex oxide – alkali halide interface partially change their effective oxidation state from 3+ down to 2+. An uncompensated magnetic moment formed due to a difference between the magnetic moments attributed to the $Fe^{3+}$ and $Fe^{2+}$ ions within the antiferromagnetic matrix leads to a notable increase of remanent magnetization of the composites. Remanent magnetization estimated for the composites does not significantly changes with an increase of KBr and it remains nearly stable for the composites with KBr fraction of about (20 – 50) %, which can be explained assuming that the chemical processes occur only in the thin sub-surface layer of the crystallites of the ferrite component. An average size of the ferrite crystallites is about 100 nm as confirmed by scanning electron microscopy measurements (see **Fig. 1**), that causes the presence of a concentration threshold (~ 20%) affecting the chemical processes and the composites having smaller KBr fraction does not show significant difference in magnetic and transport properties as compared to the ferrite compounds (see **Supplement, Fig. S4**). The origin of this effect can be a magnetic percolation that's threshold is passed at ~20% of KBr content. This idea is verified by the isothermal magnetization curves obtained for composites with the KBr phase ratios 15% and 10% which have revealed almost absence of the remanent magnetization, while distinct opened hysteresis loop is observed for the composite having 20% of KBr content, so there is really some magnetic percolation threshold. Notably, that the single-crystalline $BiFeO_3$ powder synthesized by using molten KCl-KBr salt at 750 °C showed a very weak ferrimagnetic nature at low magnetic field [32].

In order to get more information concerning the effect of KBr matrix we have measured the electroresistivity of the nanocomposites $(BLFO)_x–(KBr)_{1-x}$ and nanograined ceramics BLFO produced by the sol-gel method. Current-voltage (I-V) characteristics of all these samples are quasi-linear (see **Supplement, Fig. S4**), and the values of the electro-resistivity determined from I-V characteristics are listed in **Table I.** The I(V) curves and resistivity values in the **Table I** demonstrate the pronounced decrease in electroresistivity with the increase of KBr content. Since we can measure only electronic conductance from BLFO (not ionic contribution from the KBr phase), the changes in resistivity can be associated with the formation of $Fe^{2+}$ ions and this correlates with an increased magnetization because of uncompensated magnetic moments in antiferromagnetic interactions $Fe^{2+}$ - O - $Fe^{3+}$.

**Table I.** Electro-resistivity the nanocomposite $_x(BLFO)−_{(1-x)}KBr$

| Resistivity (in $10^9$ Ω/cm) | Composite $_xBi_{0.9}La_{0.1}FeO_3−_{(1-x)}KBr$ | | | | |
|---|---|---|---|---|---|
| | x=1 | x=0.9 | x=0.7 | x=0.5 | x=0 |
| | >10 | 5 | 2 | 0.1 | <0.01 |

## III. THEORETICAL EXPLANATION
### A. Discussion of the possible mechanisms

Different physical mechanisms (analyzed in the next subsection) can be responsible for the appearance of room-temperature weak ferromagnetism of $Bi_{1-x}La_xFeO_3$ nanosized inclusions. Recently we constructed a comprehensive Landau-Ginzburg-Devonshire (**LGD**) thermodynamic potential and using it modelled the phase diagram of pristine BFO [33]. The role of the AFD, rotomagnetic (**RM**), and rotoelectric (**RE**) couplings was established in Ref.[34]. This in complex allows reconstructing the phase diagram of BFO and long-range order parameter distributions including the temperature stability of the AFM, FE, and AFD phases, as well as prediction of novel intermediate structural phases.

Actually, $Bi_{1-x}La_xFeO_3$ can exhibit weak ferromagnetic properties near the surface of micro- and especially nanoparticles via structural distortions unlocing the cycloid modulated spin structure and resulting in $Fe^{3+}$ spins canting due to non-vanishing Dzyaloshinskii-Moriya (DM) interaction [17, 35, 36, 37, 38, 39]. Different physical mechanisms can be responsible for the appearance of room-temperature ferromagnetism of ferroelectric $Bi_{1-x}La_xFeO_3$ inclusions, while $Bi_{1-x}La_xFeO_3$ is FE and purely AFM in the bulk [3]. Such mechanisms, in particular, are surface piezomagnetic effect existing near the surface of any antiferromagnetic due to the absence of spatial inversion center at the surface [40], flexo-magnetoelectric [41] and linear antiferrodistortive-antiferromagnetic [42] couplings, Vegard effect (so called "chemical pressure") [43, 44] as well as omnipresent magnetic defects and/or impurities accumulation at the surface due to the strong lowering of their formation energy at the surface [45, 46].

According to this, different physical mechanisms should be considered to explain the unexpected experimental result for 0.5(BLFO)–0.5(KBr) and 0.7(BLFO)–0.3(KBr) nanocomposites (the model is shown in **Fig 3**), but only those among them, which are highly sensitive to the interaction of antiferromagnetic BLFO with ionic-conductor KBr, can be relevant.

Below we will try to prove that the appropriate mechanism turned out to be **ferro-magneto-ionic coupling**, that is a natural extension of ferro-ionic and anti-ferroionic couplings [47, 48, 49, 50] revealed earlier in ultra-thin ferroelectric films exposed to ionic exchange with ambient media (see original papers [51, 52], review [53] and Refs therein). Following Stephenson and Highland (**SH**) model [51, 52] the screening by ions is electrically coupled to the electrochemical processes at the ferroelectric surface, and thus, the stabilization of ferroelectric state in ultrathin $PbTiO_3$ films occurs due to the chemical switching [54, 55, 56]. The analysis [48, 50] leads to the elucidation of the ferro-ionic, anti-ferroic and electret-like non-ferroelectric states, which are the result of nonlinear electrostatic interaction between the ferroelectric polarization and absorbed ions.

In accordance with experiments, the critical size of ferroelectricity and antiferromagnetism disappearance in $BiFeO_3$ nanoparticles is very small – about 10 nm [57]. So that the "core" of 100-nm $Bi_{0.9}La_{0.1}FeO_3$ particle for sure should be ferroelectric (**FE**) and anti-ferromagnetic (**AFM**). The ferromagnetism can be induced by the ferro-ionic exchange only in the vicinity of surface [58], namely in a thin "shell" of thickness 2 – 5 nm (see **Fig. 3**).

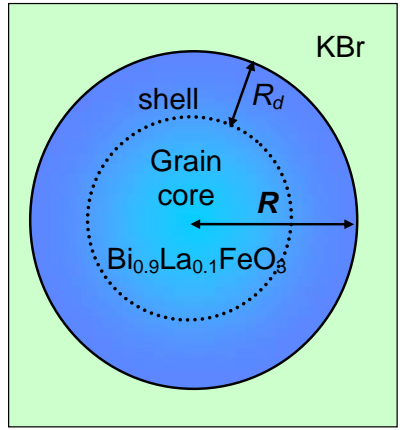

**FIGURE 3. (a)** Schematics of the spherical grain with radius *R* covered by the shell of thickness $R_d$, where the defects are accumulated. Adapted from Ref. [34].

In accordance with experimental results presented above (see **Fig. 2**), one can assume that in $(BLFO)_x$–$(KBr)_{1-x}$ nanocomposites the concentration of the "frustrated" spins in the shell of $Bi_{0.9}La_{0.1}FeO_3$ nanoparticle first increases and then saturates with the increasing fraction of KBr, 1-x. Thus, the idea of **ferro-magneto-ionic coupling** is that the surface of ferroelectric $Bi_{0.9}La_{0.1}FeO_3$ is in the dynamic direct and indirect electron exchange with ionic conductor KBr. The role of the exchanges is twofold. The first are the electronic states near the ferrite surface, which can lead to a rather weak ferromagnetic exchange due to the linear ME effect at the surface (as well as to DM-interaction, and/or geometric frustration). The mechanism will not be considered below.

The second is due to the fact that the interaction with nonmagnetic KBr matrix, via the rotostriction coupling [34], can induce the magnetic dipoles in the vicinity of surface of the $Bi_{0.9}La_{0.1}FeO_3$ inclusions, similar to the appearance of defect-driven magnetism in thin films of nonmagnetic oxides $HfO_2$, $TiO_2$ and $SnO_2$ [59, 60, 61, 62, 63, 64, 65, 66], where different vacancies can lead to the appearance of $d^0$ magnetism (but it is still under debate if there oxygen vacancies contribute only or cation vacancies also contribute). At 25% concentration of ferromagnetic spins at the ferrite surface and under the surface, magnetic percolation between them in the particle "shell" cannot be excluded. In this case, the appearance of a loop with a relatively large coercive field occurs due to the mechanism of blocking the elementary spins.

### B. Theoretical modeling

To quantify the above idea, we use the thermodynamic potential of LGD-type that describes AFM, FE, and AFD properties of BFO nanoceramics, including the RM, RE and ME biquadratic couplings, and the AFD, FE, AFM contributions, as well as elastic energy in the form [34].

Gibbs free energy of the nanoparticle (**NP**) has the following form:

$$G = G_M + G_{MS},\qquad(1)$$

where $G_M$ and $G_{MS}$ are the magnetic and magneto-elastic (magnetostriction and rotosriction) contributions to the NP energy, respectively.

The magnetic part of the free energy $G_M$ describing the AFM-FM order-disorder type transition in the NP has the form:

$$G_M = V\left(\frac{J}{2N_d}\langle l_i\rangle^2 + \frac{J_{nl}}{4N_d}\langle l_i\rangle^4 + \frac{k_B T}{2N_d}\left[(1+\langle l_i\rangle)\ln(1+\langle l_i\rangle) + (1-\langle l_i\rangle)\ln(1-\langle l_i\rangle)\right] - M_S\langle l_i\rangle H_i\right).\qquad(2)$$

The NP average volume is $V = \frac{4\pi}{3}R^3$. The dimensionless order parameter $\langle l_i\rangle$ is a degree of the magnetic dipoles ordering in the NP ($i$=1, 2, 3). The value $\langle l_i\rangle$ is statistically averaged over the orientations of elementary magnetic dipoles $l_i$ in the NP. The concentration $N_d$ of magnetic dipoles is related with the concentration of ions at the NP surface. $J$ is the exchange constant that is positive ($J \geq 0$) for the considered case of the AFM core, and related with Neel temperature $T_N$ as $J = k_B T_N$ in the mean field approximation (Boltzmann constant is $k_B = 1.38 \times 10^{-23}$ J/K). $J_{nl}$ is the nonlinear exchange constant. The macroscopic magnetization components, $M_i = M_S\langle l_i\rangle$, are coupled with the applied magnetic field $H_i$.

The magneto-elastic contribution to the free energy (1) is:

$$G_{ME} = -\left\langle Z_{klij}\sigma_{kl}M_S^2 l_i l_j + \frac{s_{ijkl}}{2}\sigma_{ij}\sigma_{kl} + u_{ij}^W\sigma_{ij}\right\rangle.\qquad(3)$$

In Equation (3) $\sigma_{ij}$ is the elastic stress tensor, $Z_{ijkl}$ is the magnetostriction stress tensor, and $s_{ijkl}$ is the elastic compliances tensor of the AFM material. The summation is performed over all repeated indices. The bracket $\langle ...\rangle$ means the statistical averaging (summation) that is regarded equivalent to the intergation over the NP volume, $\int_V d^3r(...)$ in the ergodic case. The last term in Eq.(3) is the Vegard-type energy density

$$u_{ij}^W\sigma_{ij} = \sigma_{ij}W_{ij}(\mathbf{r})\delta N_S(\mathbf{r}) \approx z_{ijkl}W_{ij}(\mathbf{r})\delta N_S(\mathbf{r})M_S^2 l_k l_l + o(l_k^2,\delta N_S^2),\qquad(4)$$

where we denote the elastic dipole tensor (Vegard expansion) of a surface defect as $W_{ij}$ [67]. Tensor $z_{ijkl} = Z_{ijmn}c_{mnkl}$ is the magnetostriction strain tensor, $c_{mnkl}$ is the elastic stiffness. The value $\delta N_S(\mathbf{r}) \sim \sum_k \delta(\mathbf{r} - \mathbf{r}_k)$ is the random concentration of elastic defects in the NP shell (including complexes with the surface ions and related with them). The approximate equality in Eq.(4) is valid if the main magnetization-dependent part of the stress is $\sigma_{ij}^M = z_{ijkl} M_S^2 l_k l_l$ due to the magnetostriction mechanism. The function $o(l_k^2, \delta N_S^2)$ designates the small high order terms.

Allowing for the presence of the term $\sigma_{ij}^M = z_{ijkl} M_S^2 l_k l_l$, the energy (3) changes the coefficient $\frac{J}{2N_d}$ in the term $\frac{J}{2N_d}\langle l_i \rangle^2$ in Eq.(2). Thus, the substitution of elastic fields (4) into the Eq.(3) and then to the Gibbs potential Eq.(1) leads to the renormalization of the coefficient $\frac{J}{2N_d}\langle l_i \rangle^2$ in Eq.(2), namely:

$$\frac{J}{2N_d}\langle l_i \rangle^2 \to \frac{J}{2N_d}\langle l_i \rangle^2 + \left\langle z_{kmii} M_S^2 \sum_n W_{km} \delta N_S(\mathbf{r} - \mathbf{r}_n) l_i l_i \right\rangle$$
$$\approx \left(\frac{J}{2N_d} + z_{kmii} M_S^2 \left\langle \sum_n W_{km} \delta N_S(\mathbf{r} - \mathbf{r}_n) \right\rangle\right) \langle l_i \rangle^2 \quad (5)$$

The statistical averaging over defect distribution in expression (5) gives:

$$\left\langle \sum_n W_{ij} \delta N_S(\mathbf{r} - \mathbf{r}_n) \right\rangle \cong W_{ij} \frac{1}{V} \int_V N_S(\mathbf{r}) d\mathbf{r} . \quad (6)$$

The summation in Eq.(6) is performed over defect sites and the averaging of the function $\delta N_S(\mathbf{r} - \mathbf{r}_l)$ in the equation leads to the integration over the shell region, where the defects are accumulated. For the clarity we assume that the distribution function of defects $N_S(\mathbf{r})$ depends on the distances from the NP surface $r = R$ (as the strongest inhomogeneity), and has exponential decay far from the surface [see **Fig. 3**]:

$$N_S(\mathbf{r}) = N_S(x) \exp\left(-\frac{R-r}{R_d}\right), \quad 0 < r < R \quad (7)$$

Here $R_d \ll R$ is the decay length of defect concentration under the surface. The amplitude $N_S(x)$ depends on the KBr fraction (1-$x$) of NP surrounding, and preparation conditions of the composite. Indeed the ionic surrounding affects the defect formation energies, in accordance with e.g. Stephenson-Highland ionic adsorption model [51, 68]. From Eq.(7), the average concentration of defects is

$$\overline{N}_S(x,R) = \frac{1}{V}\int_V N_S(\mathbf{r})d\mathbf{r} \approx 3N_S(x)\left(\frac{R_d}{R}\right). \tag{8}$$

The approximate equality is valid at $R_d \ll R$.

The temperature of the possible FM transition induced by surface ions can be defined from the expansion of the free energy (2) in the mean-field approximation, namely from the expansion up to quadratic terms of the expression

$$\begin{aligned}&\left[\frac{J}{2N_d} + z_{kmii}W_{km}\overline{N}_S M_S^2\right]\langle l_i\rangle^2 + \frac{J_{nl}}{4N_d}\langle l_i\rangle^4 \\ &+ \frac{k_B T}{2N_d}\left[(1+\langle l_i\rangle)\ln(1+\langle l_i\rangle) + (1-\langle l_i\rangle)\ln(1-\langle l_i\rangle)\right] \approx \frac{\alpha_i}{2}\langle l_i\rangle^2 + \frac{\beta}{4}\langle l_i\rangle^4 + ..O\langle l_i\rangle^5\end{aligned}, \tag{9a}$$

The coefficients α and β have the following form:

$$\alpha_i(x,R,T) = \frac{J}{N_d} + 2z_{kmii}W_{km}\overline{N}_S(x,R)M_S^2 + \frac{k_B T}{N_d}, \quad \beta(T) = \left(\frac{k_B T}{3} - J_{nl}\right)\frac{1}{N_d}. \tag{9b}$$

The critical temperature $T_{cr}^i(x,R)$ satisfies the equation $\alpha_i(x,R,T) = 0$ and acquires the form:

$$T_{cr}^i(x,R) = -\frac{J + 2z_{kmii}W_{km}\overline{N}_S(x,R)N_d M_S^2}{k_B}. \tag{10}$$

Since $J > 0$, the product $z_{kmii}W_{km}\overline{N}_S(x,R)N_d$ should be negative in order to make $T_{cr}(x,R)$ positive under the condition $J + 2z_{kmii}W_{km}\overline{N}_S(x,R)N_d M_S^2 < 0$. Hereinafter we assume that the Vegard tensor is isotropic and diagonal, i.e. $W_{ij} = W\delta_{ij}$, and magnetostriction tensor symmetry is cubic, i.e. $z_{1111} = z_{2222} = z_{2222} \equiv z$. So that $z_{kmii}W_{km} \equiv z\cdot W$.

For a semi-quantitative description of the magnetization curves observed in our experiments [see **Fig. 2**] we model the dynamics of magnetization dependence on the quasi-static magnetic field from the relaxation time-dependent equation

$$\Gamma\frac{dl}{dt} = \left(\frac{J}{N_d} + 2zWM_S^2\overline{N}_S\right)l + \frac{J_{nl}}{N_d}l^3 + \frac{k_B T}{2N_d}\ln\left(\frac{1+l}{1-l}\right) - HM_S. \tag{11}$$

Hereinafter $\Gamma$ is the relaxation coefficient. In the static case the right-hand side of Eq.(11) can be expanded in *l*-series as $\alpha l + \beta l^3 - HM_S$. The values of the remanent magnetization and "intrinsic" thermodynamic coercive field $H_c$ can be estimated from the expansion as:

$$M_r(x, R, T) \approx \sqrt{-\frac{\alpha}{\beta}}, \qquad H_C(T_r, h) = \alpha\sqrt{\frac{27\alpha}{4\beta}}. \qquad (12)$$

The dependence of magnetization of $(BLFO_3)_x$–$(KBr)_{1-x}$ nanocomposite on quasi-static magnetic field was calculated from Eq.(11) at RT. Results are presented in **Fig. 4(a)**. Different curves correspond to the gradual decrease of the BLFO fraction x = 1, 0.9, 0.8, 0.7, 0.5, 0.25, 0.1 and 0.01 (see labels near the curves). Two insets **Fig. 4(b)** and **4(c)** illustrate the dependence of the remanent magnetization ($M_r$), maximal in-field magnetization ($M_m$) and coercive field ($H_c$) on the fraction 1-x of KBr.

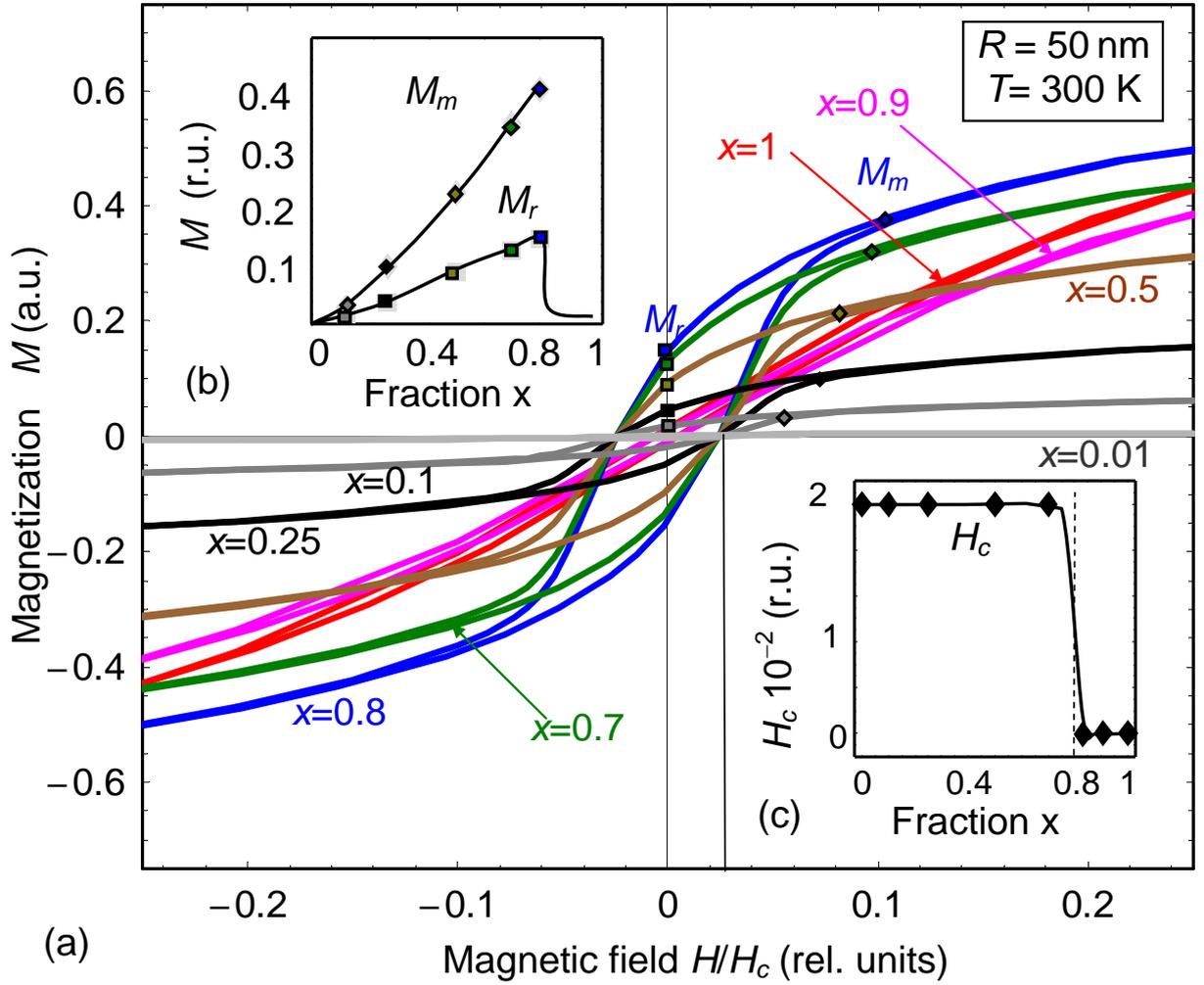

**FIGURE 4. (a)** Magnetization dependence on quasi-static magnetic field calculated for $(Bi_{0.9}La_{0.1}FeO_3)_x$ – $(KBr)_{1-x}$ nanocomposite at RT. Different curves (1 – 7) correspond to the gradual decrease of the $Bi_{0.9}La_{0.1}FeO_3$ fraction x = 1, 0.9, 0.8, 0.7, 0.5, 0.25, 0.1 and 0.01 (see labels near the curves). Two insets **(b)** and **(c)** show the dependence of the remanent magnetization ($M_r$), maximal in-field magnetization ($M_m$) and coercive field ($H_c$) on the fraction x of KBr. The average radius R of $Bi_{0.9}La_{0.1}FeO_3$ nanoparticles was 50 nm,

parameters $J/k_B = -650$ K, $z = 5 \times 10^{-3}$ magnetic units, $\overline{N}_S = (1.4 - 1.6) \times 10^{23}$ m$^{-3}$ for the curves (1 – 7), respectively, $W = -10$ Å$^3$, and $R_d = 2$ nm.

There are evident qualitative similarities (such as the loop order and shape changes with $x$ increase) between **Fig. 4** and **Fig. 2**. However they differ quantitatively. In particular, the calculated concentration dependence of the remanent magnetization and the saturation law at high magnetic fields significantly differs from the experiment. It appears that the experimental data shown in **Fig.2** can be rigorously fitted (with not more than several % error) by the function

$$M(H) = M_0 \tanh\left(\frac{H - H_S}{\Delta}\right) - \chi H - M_H \qquad (13)$$

Here $M$ is the magnetization, $H$ is the applied magnetic field. The fitting parameters $M_0$, $H_S$, $\Delta$, $\chi$ and $M_H$ depend on BLFO concentration $x$ and are listed in **Table S-I** in the **Supplement.** They are different for the "up" and "down" parts of the loops, and the small values of $M_H$ can be regarded as the error. Notably that Eq.(13) contains a small linear term $\chi H$ not included in the theoretical model (11), which magnitude depends on x% of BLFO. The presence of the linear term may be associated with the net response of the Bi$_{0.9}$La$_{0.1}$FeO$_3$ ceramics, which magnetization quasi-linearly increases with $H$. The presence of the nonlinear term $M_0 \tanh\left(\frac{H - H_S}{\Delta}\right)$ corroborates the validity of the Ising-type order-disorder Eq. (11) used for the qualitative understanding of the experimental results. Actually, the approximate analytical solution of the static equation (11) is $l = \tanh\left(\frac{H M_S N_d}{k_B T}\right)$ that is rigorous at negligibly small contributions of the linear and nonlinear terms $\left(\frac{J}{N_d} + 2 z_{ijkk} W_{ij} \overline{N}_S M_S^2\right) l$ and $\frac{J_{nl}}{N_d} l^3$, respectively. The nonlinear contribution to the magnetostatic response is attributed with ferro-magneto-ionic coupling between BLFO and KBr.

### IV. CONCLUSION

We studied magnetostatic response of the composites consisting of nanosized ferrite Bi$_{0.9}$La$_{0.1}$FeO$_3$ conjugated with fine grinded ionic conducting KBr powder. When the fraction of KBr is rather small (less than 15 %) the magnetic response of the composite is very weak and similar to that observed for the compound Bi$_{0.9}$La$_{0.1}$FeO$_3$; pure KBr matrix without Bi$_{1-x}$La$_x$FeO$_3$ has no magnetic response at all as anticipated. However, when the fraction of KBr increases above 15%, the magnetic response of the composite changes substantially and the field dependence of magnetization discloses ferromagnetic-like hysteresis loops with the remanent magnetization about 0.14 emu/g and the coercive field about 1.8 Tesla (at RT). Nothing similar to the ferromagnetic-like hysteresis loop can be observed in $Bi_{0.9}La_{0.1}FeO_3$ ceramics, which magnetization quasi-linearly increases with magnetic field.

Different physical mechanisms were proposed to explain the unusual experimental results for nanocomposites $Bi_{0.9}La_{0.1}FeO_3$-KBr, but only those among them, which are highly sensitive to the interaction of antiferromagnetic $Bi_{0.9}La_{0.1}FeO_3$ with ionic conductor KBr, can be relevant. We have shown that the ferromagnetic behaviour appears as a synergetic effect driven by the ferro-magneto-ionic coupling.

**Acknowledgements**

The work has received funding from the European Union's Horizon 2020 research and innovation programme under the Marie Skłodowska-Curie grant agreement No 778070, and partially supported by the National Academy of Sciences of Ukraine (project No. 0117U002612) and RFBR (according to the research project № 18-38-20020 mol_a_ved). S.V.K. was supported by the Center for Nanophase Materials Sciences, sponsored by the Division of User Facilities, Basic Energy Sciences, US Department of Energy.

**Authors' contribution.** D.V.K. conducted and supervised magnetostatic and electrophysical experiments. O.M.F. (jointly with M.C. and A.Y.) prepared the first composite samples. M.V.S. and D.V.K. prepared pristine BFO and several composite BLFO and BFLO-KBr (jointly with S.V.D., Y.G., A.P., R.S., A.L. and A.K.). A.N.M. generated the research idea, obtained theoretical results (jointly with O.M.F.) and wrote the manuscript draft. W.S., V.V.S., S.V.K., and N.V.M. densely worked on the results interpretation and manuscript improvement.

**SUPPLEMENT**

**SAMPLES CHARACTERIZATION**

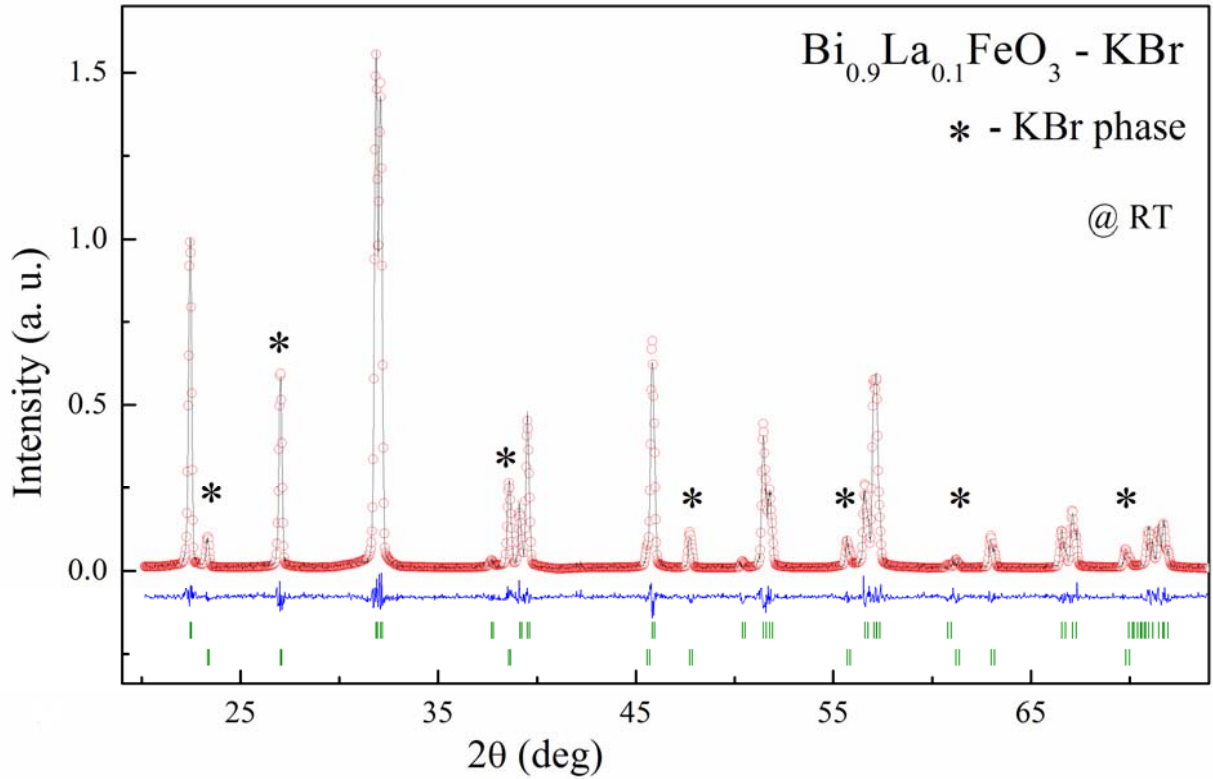

**Figure S1**. Refined XRD pattern obtained for the composite $(Bi_{0.9}La_{0.1}FeO_3)_{0.8}$–$(KBr)_{0.2}$ at room temperature (circles are experimental data, lines are calculated ones). Upper row of the Bragg reflections are associated with the ferrite phase (space group R3c), KBr phase is denoted by asterisks.

For the sake of comparison with the composite, the temperature dependencies of $Bi_{1-x}La_xFeO_3$ (x=0, 0.1 and 0.15) magnetization have been measured in field cooled (upper curves "FC" in **Fig. S2**) and zero field cooled (lower curves "ZFC" in **Fig. S2**) modes in a static magnetic field of 1kOe, and temperatures from 300 K up to 1000 K, using MPMS SQUID VSM magnetometer.

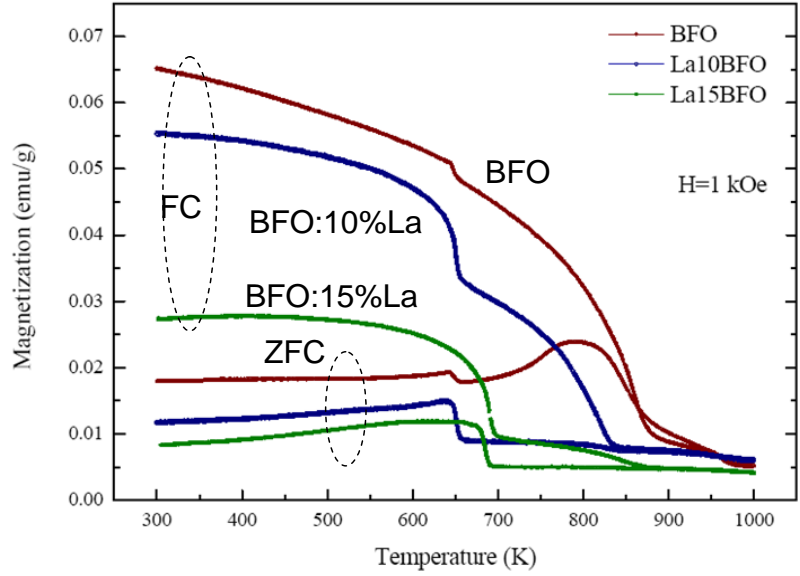

**Figure S2.** Temperature dependencies of $Bi_{1-x}La_xFeO_3$ magnetization (x=0, 0.1 and 0.15) measured in the field cooled (brown, dark blue and green upper curves "FC") and zero field cooled (brown, dark blue and green lower curves "ZFC") modes in a static magnetic field of 1kOe at temperatures 300 – 1000 K.

For the sake of comparison with the composite, we studied how the increase of La content affects on the field dependence of magnetic moment M(H) of the AFM compound $Bi_{1-x}La_xFeO_3$ for x = (0 - 0.17). Isothermal dependencies of magnetization M(H) shown in **Fig. S3** demonstrate nearly linear behavior with gradual decrease in coercivity with x decrease for lightly doped compounds up to magnetic field of 5 Tesla. M(H) dependencies measured in fields of ~ 10T (measured using magnetometer from Cryogenic Ltd.). for $BiFeO_3$ and $Bi_{0.95}La_{0.05}FeO_3$ and M(H) dependencies for heavily-doped compounds show metamagnetic transition behavior associated with a disruption of spatially modulated magnetic structure (specific for initial BFO).

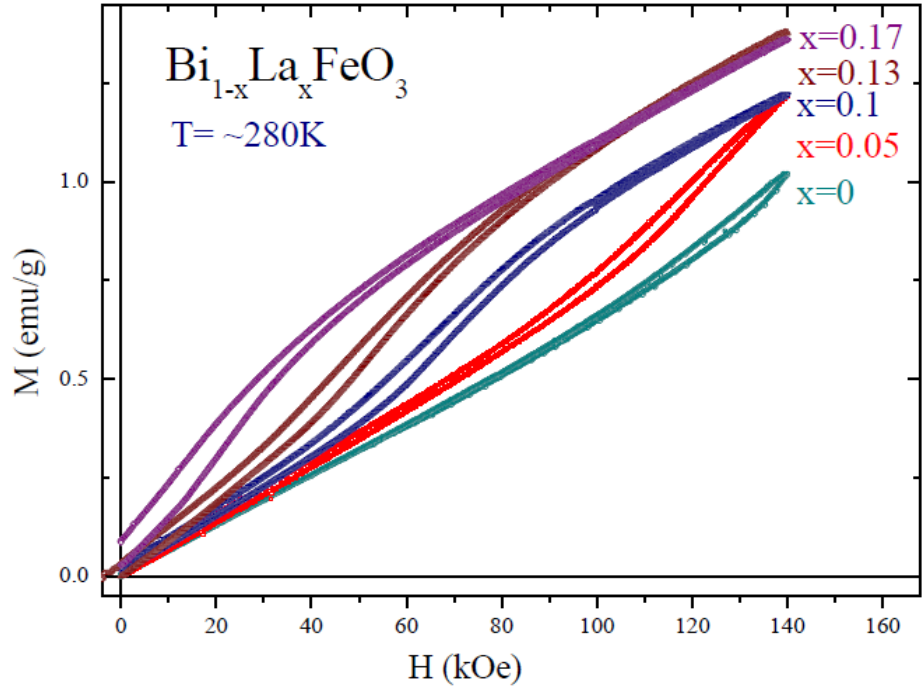

**Figure S3.** Dependencies of magnetic moment on applied magnetic field (in Tesla) measured for $Bi_{1-x}La_xFeO_3$, where x = 0, 0.05, 0.1, 0.13 and 0.17 (different curves symbols) at room temperature.

Volt-ampere characteristics of $Bi_{0.9}La_{0.1}FeO_3$ – KBr, and $Bi_{0.9}La_{0.1}FeO_3$ ceramics are shown in **Fig. S4.**

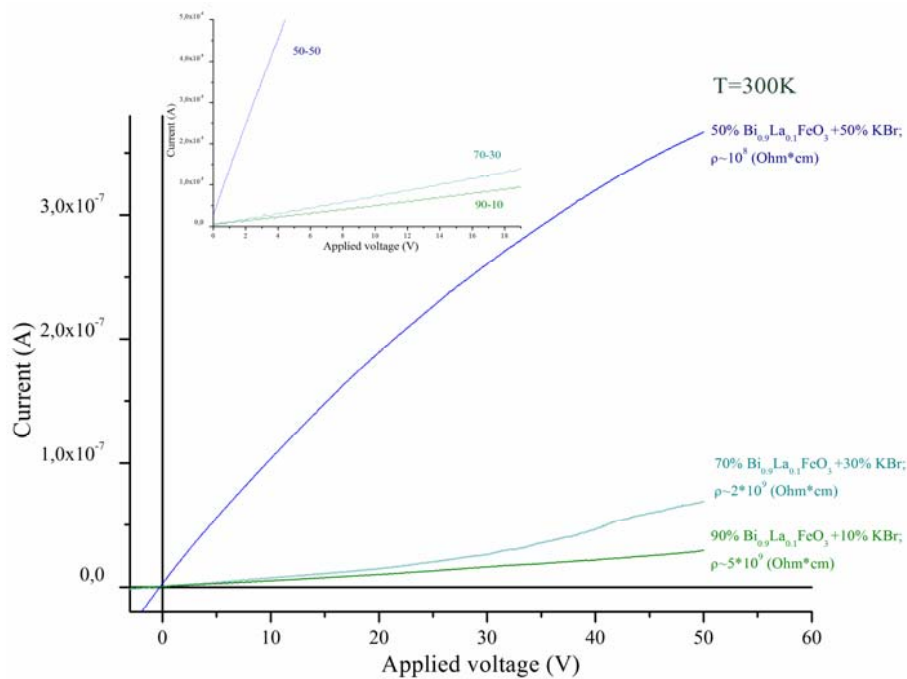

**Figure S4.** Volt-ampere characteristics of $Bi_{0.9}La_{0.1}FeO_3$ – KBr, and $Bi_{0.9}La_{0.1}FeO_3$ ceramics.

Note about heating a mixture of BiLaFeO –KBr (50-50) and checking its color, we've heated a mixture at 100C for 30min and longer and tried to find any difference in color – I'd say there is nearly no difference, then we heated at higher temperature (up to melting point of KBR (~750C) with a step of 50-100C), at elevated temperatures BiLaFeO becomes more dark, but whole powder in a crucible (not only that being in a contact with KBr), so I'd say it's not easy (if possible) to estimate this redox reaction visually. We've also made *dc* conductance measurements (pls. see attached file) (and hope to get impedance spectroscopy results in a couple of weeks). Assuming the I(V) curves we have an increase in conductivity (the samples were in tablet form with 5mm in diameter and 1mm thickness) with KBr content. If I'm not mistaken we can see only electronic conductance from BiLaFeO (not ionic from KBr) so the changes in conductivity can be associated with a formation of $Fe^{2+}$ ions and this correlates with an increased magnetization because of uncompensated magn. moments in antiferromagnetic interactions $Fe^{2+}$ - O - $Fe^{3+}$.

**Table S-I. Fitting parameters in Eq. (13)**

| x (%) (BLFO) | $M_0$ | $H_S$ | $\Delta$ | $\chi$ | $M_H$ |
|---|---|---|---|---|---|
| **50 (up)** | -0,0917 | -2,4883 | -2,0111 | -0,0276 | -0,0036 |
| **50 (down)** | -0,0961 | 2,5029 | -2,0883 | -0,0272 | 0,0018 |
| **70 (up)** | -0,1492 | -2,7157 | -2,0508 | -0,0490 | -0,0066 |
| **70 (down)** | -0,1582 | 2,7932 | -2,1669 | -0,0484 | 0,0006 |
| **80 (up)** | -0,1565 | -3,3735 | -2,4499 | -0,0527 | -0,0055 |
| **80 (down)** | -0,1666 | 3,4416 | -2,6627 | -0,0522 | 0,0005 |
| **85 (up)** | -0,0226 | -6,6238 | -0,4442 | -0,0713 | 0,0054 |
| **85 (down)** | -0,0236 | 6,3713 | -0,00645 | -0,0713 | -0,0067 |
| **90 (up)** | -0,0303 | -6,7996 | -0,3464 | -0,0980 | 0,0070 |
| **90 (down)** | -0,0386 | 6,8332 | -0,4350 | -0,0970 | -0,0166 |
| **100 (up)** | -0,0348 | -6,3419 | -0,4935 | -0,0984 | 0,00126 |
| **100 (down)** | -0,0427 | 6,3601 | -0,6061 | -0,0975 | -0,0108 |